\documentclass[aps,pre,twocolumn,superscriptaddress]{revtex4-1}
\usepackage{graphicx}        
\usepackage{amssymb}        
\usepackage{color}           
\usepackage[dvipsnames]{xcolor}

\newcommand{\bB}{\mathbf{B}}
\newcommand{\bv}{\mathbf{v}}

\newcommand{\bzpm}{\mathbf{z^\pm}}

\begin{document}
\preprint{final}

\title{On the statistical properties of turbulent energy transfer rate in the inner Heliosphere}

\author{Luca Sorriso-Valvo}
\email[]{lucasorriso@gmail.com}
\affiliation{CNR-Nanotec, U.O.S. di Rende, ponte P. Bucci, cubo 31C, 87036 Rende (CS), Italy}
\author{Francesco Carbone}
\affiliation{CNR-Institute of Atmospheric Pollution Research, Division of Rende, UNICAL-Polifunzionale, 87036 Rende (CS), Italy}
\author{Silvia Perri}
\affiliation{Dipartimento di Fisica, Universit\`{a} della Calabria, I-87036, Rende (CS), Italy}
\author{Antonella Greco}
\affiliation{Dipartimento di Fisica, Universit\`{a} della Calabria, I-87036, Rende (CS), Italy}
\author{Raffaele Marino}
\affiliation{Laboratoire de M\'ecanique des Fluides et d'Acoustique, CNRS, \'Ecole Centrale de Lyon, Universit\'e de Lyon, 69134 \'Ecully, France}
\author{Roberto Bruno}
\affiliation{IAPS-INAF, via Fosso del Cavaliere 100, I-00133 Roma Italy}

\begin{abstract}
The transfer of energy from large to small scales in solar wind turbulence is an important ingredient of the longstanding question about the mechanism of the interplanetary plasma heating. Previous studies have shown that magnetohydrodynamic (MHD) turbulence is statistically compatible with the observed solar wind heating as it expands in the heliosphere. However, in order to understand which processes contribute to the plasma heating, it is necessary to have a local description of the energy flux across scales. To this aim, it is customary to use indicators such as the magnetic field partial variance of increments (PVI), which is associated with the local, relative, scale-dependent magnetic energy. A more complete evaluation of the energy transfer should also include other terms, related to velocity and cross-helicity. This is achieved here by introducing a proxy for the local, scale dependent turbulent energy transfer rate $\epsilon_{\Delta t}(t)$, based on the third-order moment scaling law for MHD turbulence. Data from {\it Helios 2} are used to determine the statistical properties of such a proxy in comparison with the magnetic and velocity fields PVI, and the correlation with local solar wind heating is computed. PVI and $\epsilon_{\Delta t}(t)$ are generally well correlated, however $\epsilon_{\Delta t}(t)$ is a very sensitive proxy that can exhibit large amplitude values, both positive and negative, even for low amplitude peaks in the PVI. Furthermore, $\epsilon_{\Delta t}(t)$ is very well correlated with local increases of temperature when large amplitude bursts of energy transfer are localized, thus suggesting an important role played by this proxy in the study of plasma energy dissipation.
\end{abstract}



\maketitle

\section{Introduction}
\label{Section:Intro}
The solar wind is the most important example of natural plasma turbulence that can be probed using satellite-born instrumentation~\citep{tumarsch,living}. This means space missions dedicated to {\it in-situ} measurements of plasma parameters and electromagnetic fields provide a unique chance to obtain direct experimental observation of the turbulent dynamics of space plasmas. 
Solar wind plasma has very low density (of the order of a few particle per cubic centimeter) and high temperature (of the order of $10^5$ K), and is embedded in a radially decreasing background magnetic field of the order of $B_{SW}\sim 5$ nT near the Earth, resulting in a weakly collisional, magnetized flow. 
Strong acceleration mechanisms push the wind away from the Sun to a typical speed $V_{SW} \sim 350$--$750$ km s$^{-1}$, making the wind supersonic and superalfv\'enic. Observations also show that the temperature decreases with the distance from the Sun more slowly than expected for an adiabatically expanding plasma~\citep{schwenn,freeman,goldstein}. Understanding the heating mechanism providing the non-adiabatic cooling of the expanding solar wind is a long-standing open question in astrophysics~\citep{Matthaeus94,Richardson95}. One of the possible sources of heat is the dissipation of the kinetic and magnetic energy available in the form of large scale fluctuations, which can be traced back to the Sun and the solar corona. This requires that the energy is transported from large scale to smaller scales by a turbulent cascade, where kinetic plasma processes can convert it into particle heating.
Observations of magnetic and velocity power spectral density have shown that solar wind fluctuations follow the typical Kolmogorov-like power-law energy decay~\citep{k41,frisch,marschtu,living}, in a range between the typical correlation length (corresponding to a few hours)~\citep{correlation}, and the typical scale where kinetic processes arise (a few seconds)~\citep{leamon}. This gives about three decades of inertial range where nonlinear energy transfer occurs, resulting in high Reynolds number turbulence~\citep{sorrisovalvo_scales1,weygand,sorrisovalvo_scales2}. Turbulence is therefore the major ingredient to make a connection between the large-scale fluctuations and the small-scale microphysics processes~\citep{Alexandrova2013}.

While at spatial scales smaller than the typical ion scales, kinetic processes must be included in the dynamics, at larger scales the magnetohydrodynamics (MHD) framework~\citep{biskamp} is a good approximation to describe the turbulent motion of the solar wind plasma. 
The properties of solar wind turbulence have been studied for more than forty years using both experimental data and, more recently, numerical simulations.
It is now understood that the spectral properties of the magnetic and velocity fluctuations depend on the wind speed, distance from the Sun, solar activity, correlation between velocity and magnetic field, and other local plasma parameters, making the solar wind a complex environment with high variability \citep{living}. Solar wind turbulence is also characterized by anisotropy and intermittency, which have been deeply  studied in the past \citep{Bavassano82,marschtu}. Intermittency, in particular, is related to the appearance of small-scale structures typical of turbulence. In the solar wind, these are mostly current sheets, magnetic discontinuities, vorticity structures or similar features, and are usually identified using field increments or wavelet-based detection techniques~\citep{veltri,brunoLIM,greco,Greco14,Zhdankin}. One example of identification technique recently introduced is the partial variance of increments (PVI), based on the evaluation of the intensity of the field gradients at a given position and scale~\citep{greco,Greco14}.
Recent investigations have confirmed that the intermittent structures are associated with enhanced plasma heating. Both ions~\citep{osman,tessein} and electrons~\citep{Chasapis2015} display energization in the proximity of the most intense current sheets. This has also been confirmed through the analysis of numerical simulations of the Vlasov-Maxwell equations~\citep{servidio2012}. The processes responsible for such heating may involve magnetic reconnection, plasma instabilities and enhancement of collisions, and are still poorly understood~\citep{chen}.
In this work, a different data analysis technique is proposed to identify the regions of space that are carrying energy towards the small scales, in order to understand the link between the presence of strong turbulent fluctuations and a local enhancement of temperature. 
In analogy with the von Karman-Howart law for Navier-Stokes turbulence, the MHD turbulent energy flux across the scales is regulated by a relation, often referred to as the Politano-Pouquet law (PP)~\citep{pp98}. This is a statistical prescription for the scaling law of the mixed third-order moment of the Elsasser fields increments, and is obtained directly from the MHD equations under the assumptions of stationarity, isotropy, incompressibility and vanishing dissipation coefficients (i.e. within the inertial range). 
Although some of the above assumptions are only marginally satisfied in the solar wind, the validity of the PP law has been successfully verified in numerical simulations~\citep{sorriso2002} and in the solar wind~\citep{mcbride2005,prl,apjl,mcbride2008,marino_12}. 
Subsequently, a variety of extensions of the PP law to more complex, realistic systems have been introduced, where the approximations of incompressibility~\citep{prlcomp,GaltierCOMPR} and isotropy~\citep{anisotropyCHUCK,osman2011} have been relaxed, and in some cases verified in experimental data~\citep{supratik,lina}. Effects of the solar wind expansion have also been considered~\citep{expansionGIGA,expansionPETR}, and an attempt to include the small-scale effects described by Hall MHD has been performed~\citep{GaltierHALL}. In this work, we chose not to take into account any of the above modifications, in order to provide a first-order estimation of the energy transfer rate. The evaluation of the contributions emerging when approximations are relaxed represents an interesting possible improvement that we leave to a future work. 

The basic version of the PP law reads 
\begin{equation}
    Y^\pm({\Delta t}) = \left \langle |\Delta z^\pm_{\Delta t}(t)|^2\, \Delta z^\mp_{||,\Delta t}(t)\right\rangle = -\frac{4}{3} \,\langle\epsilon^\pm\rangle{\Delta t} \langle v \rangle \; .
\label{yaglom}
\end{equation}
Here $\Delta\psi_{\Delta t}=\psi(t+\Delta t)-\psi(t)$ indicates the increment of a generic field $\psi$ across a temporal scale $\Delta t$, and the subscript $||$ indicates the longitudinal component, i.e. parallel to the bulk speed in solar wind time series; $\bzpm=\bv\pm\bB/\sqrt{4\pi\rho}$ are the Elsasser variables that couple the solar wind velocity $\bv$ and the magnetic field $\bB$, transformed in velocity units using the solar wind density $\rho$; $Y^\pm({\Delta t})$ are the mixed third-order moments, and $\langle\epsilon^\pm\rangle$ is the mean energy transfer rate, estimated over the whole domain.
In order to study spacecraft time series, all spatial scales $\ell$ were customarily transformed in the time lags $\Delta t = \ell / |\langle \bv \rangle|$ through the bulk flow speed $\langle \bv \rangle$ averaged over the entire data set. This is allowed by the Taylor hypothesis~\citep{taylor}, which is robustly valid for solar wind fluctuations in the inertial range~\citep{perriT}. 
In the right hand side of Equation~(\ref{yaglom}), $\langle\epsilon^\pm\rangle$ is the mean energy transfer rate, estimated over the whole sample. 
The PP law in Equation~(\ref{yaglom}) thus indicates that the nonlinear transport of energy across the time scales is proportional to the time scale via the mean energy transfer rate. 

In MHD numerical simulations, the statistical properties of the local energy dissipation can be studied directly~\citep{Zhdankin}.
However, when the plasma is weakly collisional, MHD viscous and resistive dissipative terms are not defined. 
In such cases, if a sufficient scale separation exists between the inertial range and the dissipative scales, even though the dissipation mechanisms are unknown the PP law can provide an estimate of the mean energy transfer rate. This has recently been measured from solar wind data, providing values compatible with the energy necessary for the observed non-adiabatic cooling~\citep{apjl,smith,prlcomp,marino_11,coburn,supratik,lina}. 

Although the PP law is only valid in a statistical sense, dimensional considerations suggest that it may be possible to use its local values as a proxy of the local energy transfer rate~\citep{marschtu}. This proxy was recently used to validate a multifractal model of the statistical properties of the turbulent fluctuations~\citep{sorriso2015}. In this paper, it will be used as a tool to identify places where energy is being transferred towards the small scales. 

By analogy with the definition of mean energy transfer rate, we define a ``local'' pseudo-energy transfer rate proxy (LET) as: 
\begin{equation}
    \epsilon^\pm_{\Delta t} (t) = \frac{|\Delta  z^\pm_{\Delta t}(t)|^2\, \Delta z^\mp_{||,\Delta t}(t)}{{\Delta t} \langle \bv \rangle} \, ,
\label{pseudoenergy}
\end{equation}
so that the local energy transfer rate at the scale ${\Delta t}$ is computed as $\epsilon_{\Delta t}(t) = (\epsilon^+_{\Delta t} (t)+ \epsilon^-_{\Delta t} (t))/2$. At a given scale, each field increment can thus be associated with the local value of $\epsilon_{\Delta t}(t)$~\citep{marschtu,sorriso2015}. In terms of velocity and magnetic field, the LET is $\epsilon_{\Delta t}(t)\propto 2\Delta v_{||}(\Delta v^2+\Delta b^2)-4\Delta b_{||}(\Delta \mathbf{v}\cdot\Delta \mathbf{b})$, where the first term is associated with the energy advected by the velocity, and the second to the velocity-magnetic field correlations coupled to the longitudinal magnetic field.
The statistical properties of this proxy will be explored here using {\it Helios 2} measurements in the inner Heliosphere.
This article is organized as follow: in Section~\ref{Section:Data} we describe the data and the diagnostic variables used for the analysis; in Section~\ref{Section:Statistical} we study the scale-dependent statistical properties of the local energy transfer rate, and compare them with the PVI, a standard indentification tool; finally, in Section~\ref{Section:Conditioned} the correlation with the local proton temperature is studied.
%
%
%
%
%
%
%
\section{Description of data}
\label{Section:Data}
This work presents a study on {\it Helios 2} data, which have been thoroughly analysed for about 40 years, and still represent a milestone in the study of the inner heliosphere. More in particular, the data selected here are 11 intervals taken during the first 4 months of 1976, at low solar activity, while {\it {\it Helios 2}} orbit spanned between 1 AU, on day 17, to 0.29 AU on day 108. Each interval includes 2178 data points at 81 second cadence, covering about 2 days of measurements. All intervals were extracted during relatively stationary wind conditions, i.e. far from the inter-stream interaction regions, and include time series of magnetic field ${\bf B}(t)$, velocity ${\bf v}(t)$, proton number density $n_p(t)$ and proton temperature $T_p(t)$. Five samples refer to slow solar wind, with average bulk speed $V_{SW}\lesssim 450$ km/sec, and six to fast solar wind, with $V_{SW}\gtrsim 550$ km/sec. General parameters of the 11 intervals are given in Table~\ref{Table:streams}. 

\begin{table}
	\caption{List of the eleven 49-hour intervals used for this work. 
	For each interval: the day of the year 
	1976 when the sample begins, DOY; the heliocentric distance, R; the mean speed, 
	$V_{SW}$ are indicated. }
	\label{Table:streams}
	\begin{tabular}{ccc} 
		\hline
		DOY & R (AU) & $V_{SW}$ (km/sec)  \\
		\hline
		 22 & 0.98 & 676  \\
		 28 & 0.97 & 348  \\
		 32 & 0.96 & 587  \\
		 46 & 0.90 & 433  \\
		 49 & 0.88 & 643  \\
		 72 & 0.70 & 411  \\
		 75 & 0.67 & 632  \\
		 81 & 0.59 & 343  \\
		 85 & 0.54 & 543  \\
		 99 & 0.35 & 431  \\
		105 & 0.30 & 727  \\
		\hline
	\end{tabular}
\end{table}

In the next sections, we will use the {\it Helios 2} data to characterize the statistical properties of LET.
%
%
%
%
%
%
%
%
%
%
%
\section{Statistical properties of the local energy transfer rate}
\label{Section:Statistical}
\begin{figure}
  \begin{center}
  \includegraphics[width=\columnwidth]{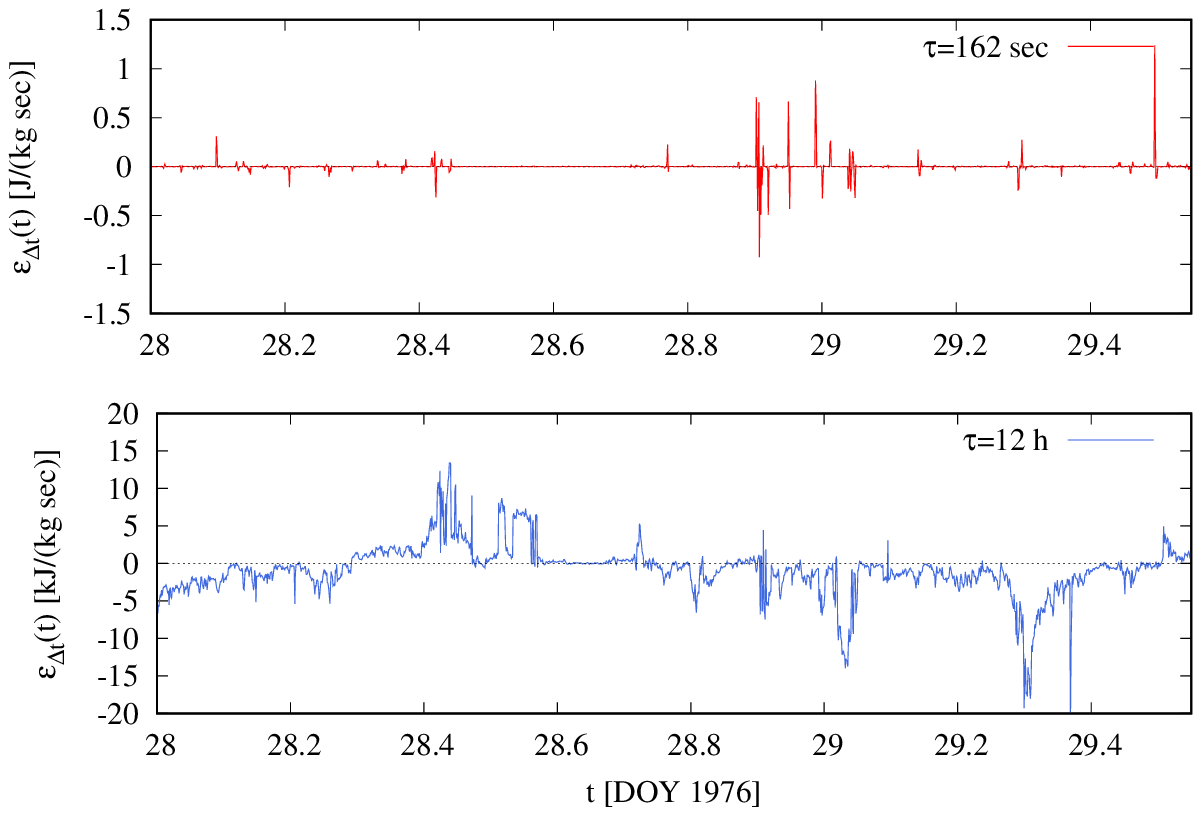}
  \includegraphics[width=\columnwidth]{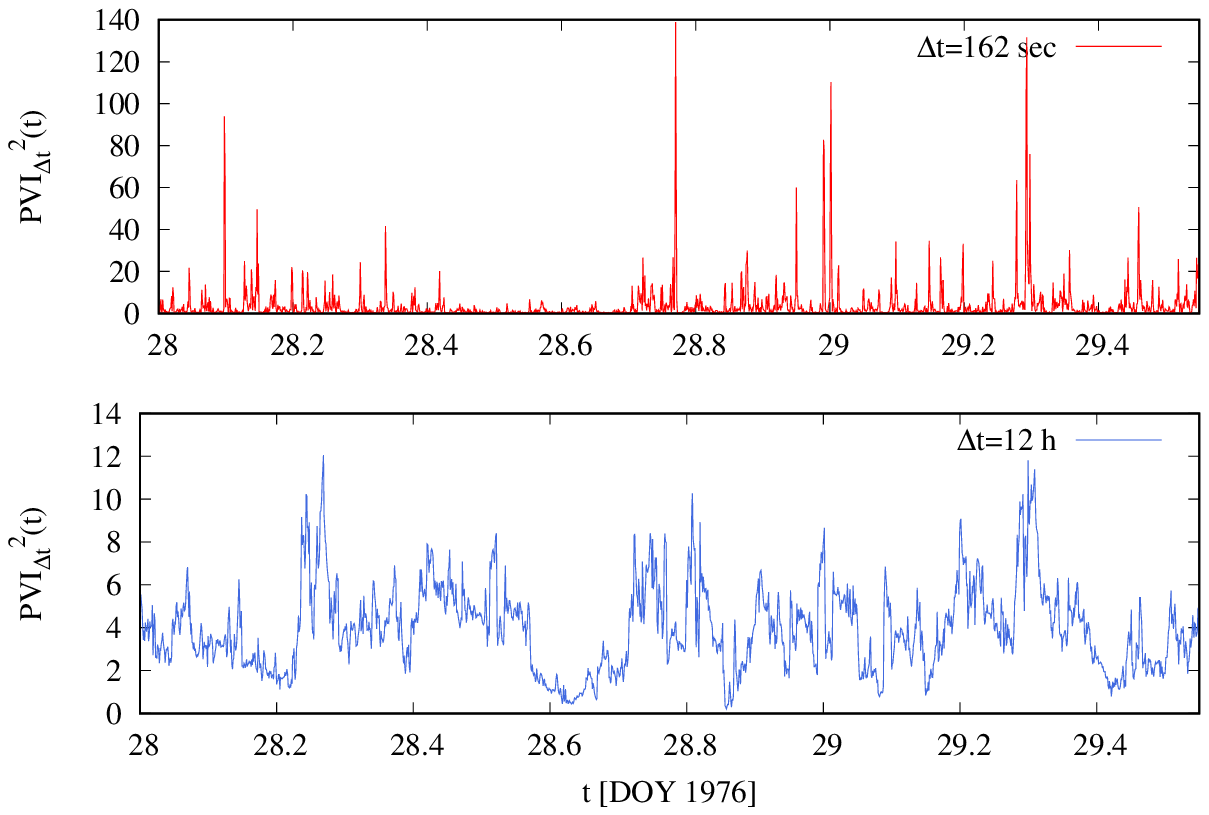}
  \hskip 304pt\includegraphics[width=0.987\columnwidth]{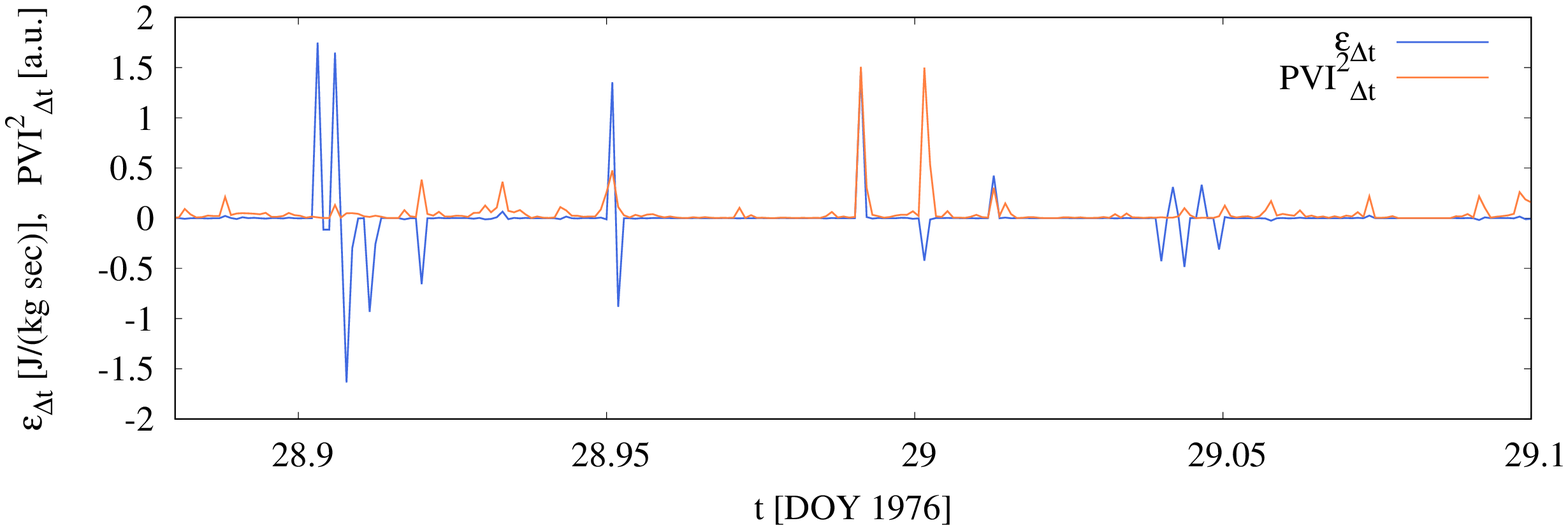}
  \end{center}
  \caption{Examples of $\epsilon_{\Delta t}(t)$ and  $PVI_{\Delta t}(t)$ 
           at two different scales $\Delta t$, for the slow wind interval of DOY 28. A magnification of both proxies 
	   at the resolution scale $\Delta t=81$ sec is shown in the {\it bottom panel}, for a shorter time interval.}
  \label{Fig:epsilon}
  \end{figure}

\begin{figure}
  \begin{center}
  \includegraphics[width=\columnwidth]{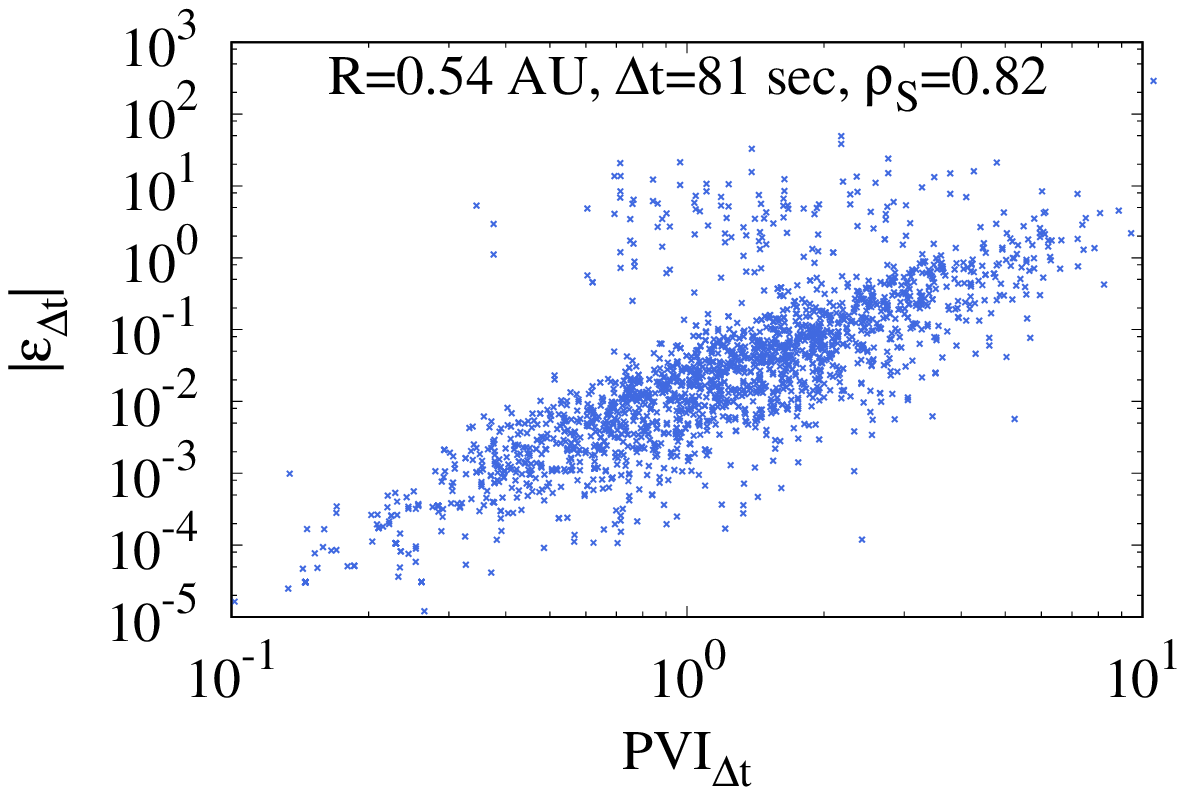}
  \includegraphics[width=\columnwidth]{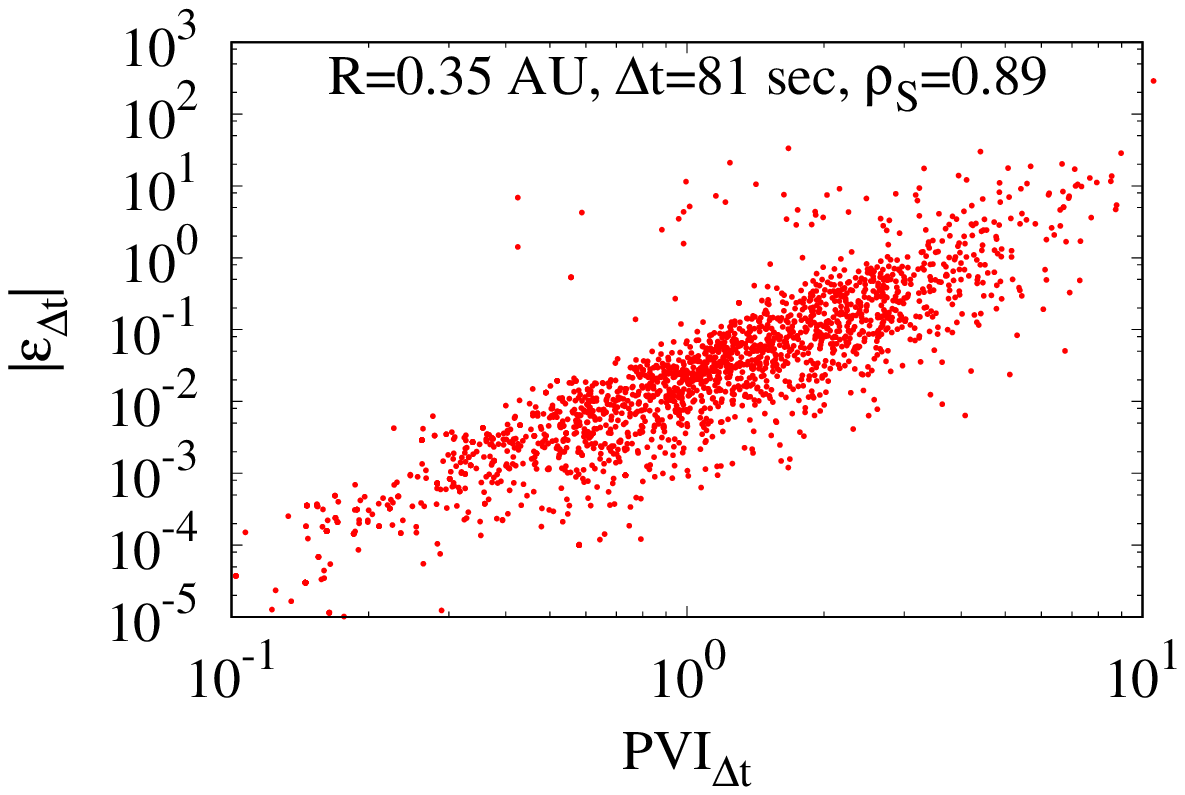}
  \end{center}
  \caption{Scatter plots of $|\epsilon_{\Delta t}(t)|$ versus $PVI_{\Delta t}(t)$
           at the smallest scale ${\Delta t}=81$ sec, for the fast interval of DOY 85 ({\it top}) 
	   and for the slow interval of DOY 99 ({\it bottom}).
           The corresponding spearman correlation coefficient $\rho_S$ are indicated.}
  \label{Fig:scatter}
  \end{figure}

The statistical properties of LET can provide quantitative information about the characteristics of the turbulent cascade occurring in the ``fluid'' range in solar wind plasma. 
For each of the samples indicated in Table~\ref{Table:streams}, we have calculated the proxy {of LET, $\epsilon_{\Delta t}(t)$,} at different time scales $\Delta t$, using equation~(\ref{pseudoenergy}) as described in Section~\ref{Section:Intro}. 
Figure~\ref{Fig:epsilon} shows one examples of $\epsilon_{\Delta t}(t)$, at two different scales (two upper panels). The scale-dependent nature of the signal is evident from the comparison of the two panels, and is the typical signature of intermittency, resulting in the strongly bursty field observed at small $\Delta t$. 
In order to compare LET with the standard PVI, we estimated the latter, including both velocity and magnetic structures contribution, as 
\begin{equation}
PVI_{\Delta t}^2(t)=\frac{\Delta \mathbf{B}(t)^2}{\Delta \mathbf{B}_{rms}^2}+\frac{\Delta \mathbf{v}(t)^2}{\Delta \mathbf{v}_{rms}^2} \; ,
\end{equation}
where $\Delta \mathbf{B}_{rms}$ and $\Delta \mathbf{v}_{rms}$ indicate the standard deviation of the magnetic field and velocity increments at the scale $\Delta t$, computed over each interval. 
The main difference between the two proxies is that, while the PVI account for the amplitude of magnetic and velocity gradients (so being an estimate of electric current and vorticity structures), the LET carries information on the magnetic and kinetic energy coupled to the advecting velocity field, and on the cross-helicity coupled to the longitudinal magnetic field. Furthermore, contrary to PVI, the LET is signed, and might contain additional information about the local direction of the energy flux.
Examples of $PVI_{\Delta t}^2(t)$ for the same interval and for the two same scales are given in Figure~\ref{Fig:epsilon}, in the third and fourth panels from the top. A similar scale-dependent burstiness is observed for the PVI, although qualitative difference is present. In particular, the LET appears burstier than PVI. 
For a direct visual comparison, in the bottom panel of Figure~\ref{Fig:epsilon} we show the two proxies at the smallest scale $\Delta t=81$ sec, in a short time interval, as to compare the correspondence of energy bursts. As evident, there are times of good correspondence between LET and PVI, but also times when only one of the two proxies has one or more peaks. 
%
%
%
%
%
%
Figure~\ref{Fig:scatter} shows scatter plots of the two variables for one fast (top panel) and one slow (bottom panel) interval, which demonstrates a good qualitative agreement between them. This is confirmed by the large associated Spearman correlation coefficient, $\rho_S\sim 0.9$. 
However, it should be noticed that the correlation is less evident when large values of $\epsilon_{\Delta t}$ and $PVI_{\Delta t}(t)$ are considered, i.e. where the energy flow is larger and at the most intense current structures. In particular, for the fast wind of DOY $85$ (top panel) there is an evident presence of points with larger $\epsilon_{\Delta t}$ and small $PVI_{\Delta t}(t)$ (the isolated population lying above the correlated points), indicating times when an enhanced energy flux does not necessarily correspond to comparably strong current sheets. This effect is still present, but less relevant in slow wind intervals, where the smaller correlations between velocity and magnetic field reduce the difference between the two proxies.
%
%
%
%
%

\begin{figure}
  \begin{center}
  \includegraphics[width=\columnwidth]{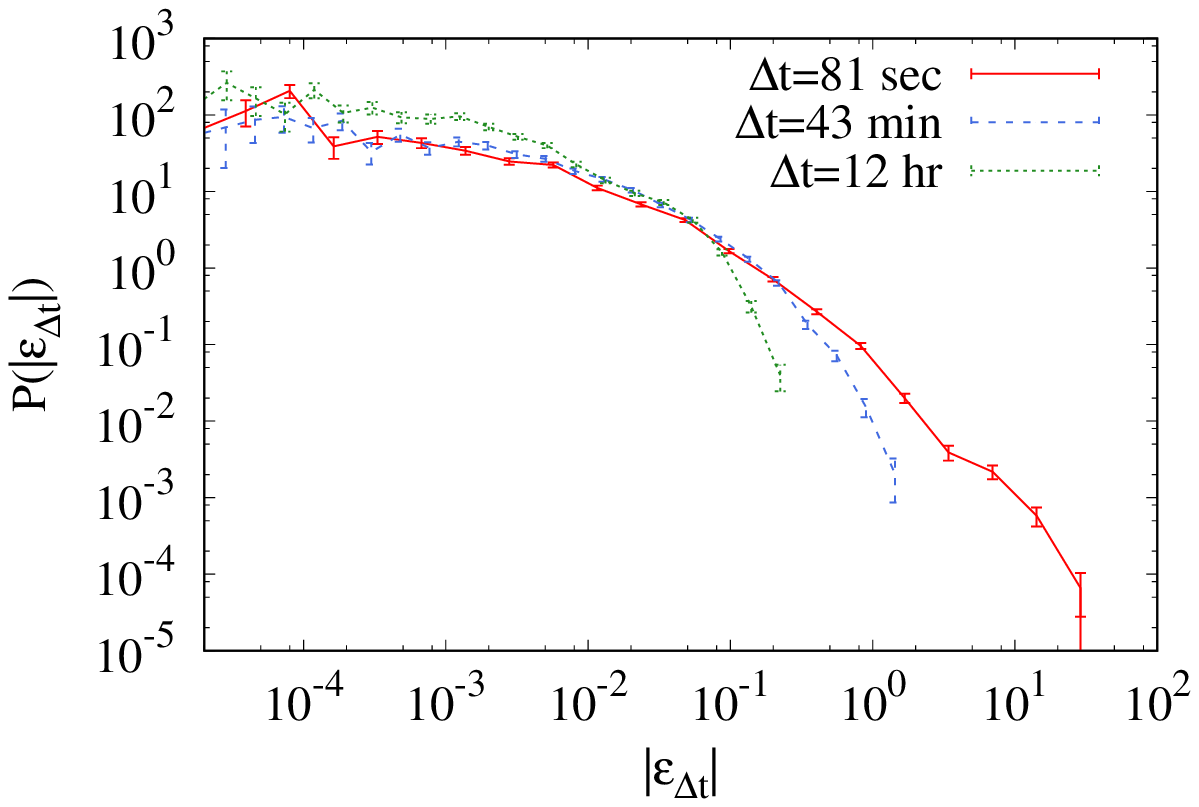}
  \includegraphics[width=\columnwidth]{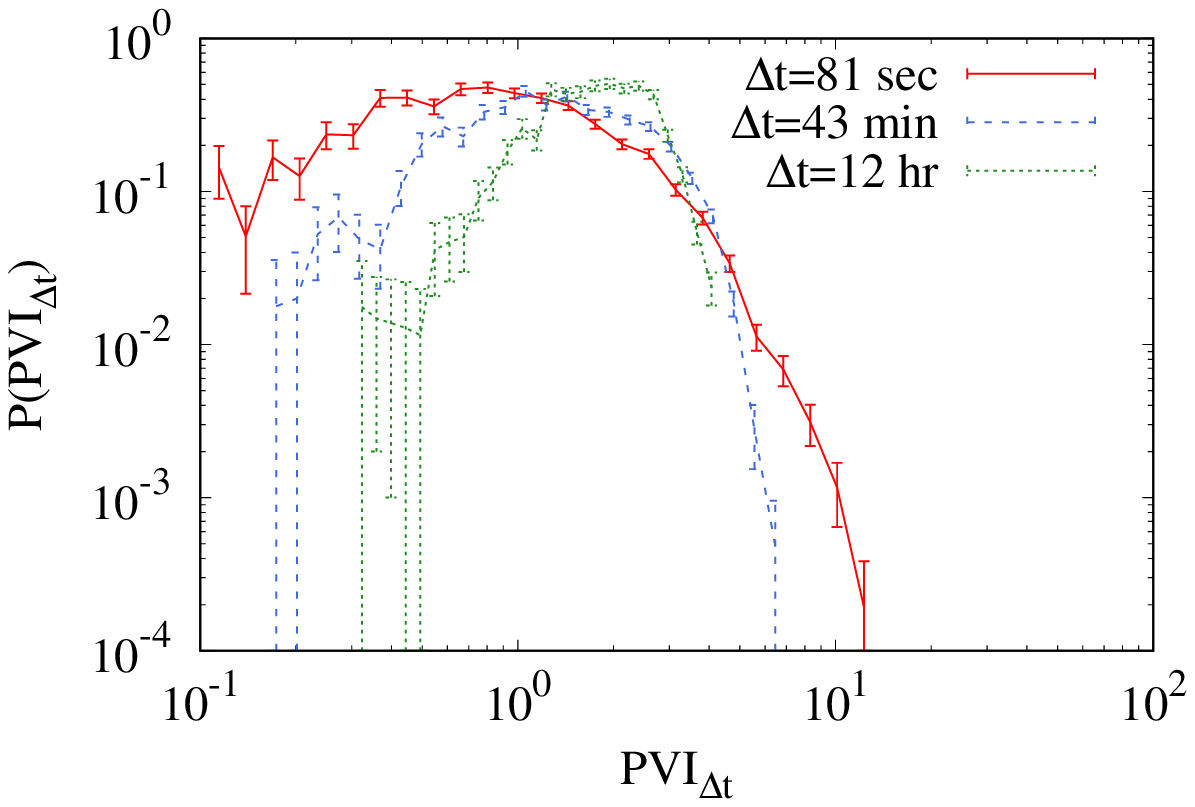}
  \end{center}
  \caption{Examples of PDFs of $|\epsilon_{\Delta t}(t)|$ ({\it top panel}) and $PVI_{\Delta t}(t)$ 
           ({\it bottom panel}) at three different scales ${\Delta t}$, for the fast wind interval of DOY 22.
	   The error bars are obtained by propagating the standard Poisson statistical uncertainity of each bin count. }
  \label{Fig:pdfs}
  \end{figure}

In order to fully describe the statistical properties of the two proxies, we compute for each interval the probability distribution function (PDF) of both variables, at different scales. For the LET, we show the distribution $P(|\epsilon_{\Delta t}(t)|)$, having verified that the positive and negative parts of the variable have very similar statistics. Figure~\ref{Fig:pdfs} illustrates the difference between the two proxies. It can be noted that at large values the functional form of the PDFs of the two proxies change. Although most of the data are in the core of the distribution (small values), the interesting large bursts represented in the right tails are particularly relevant for this work. 
Both proxies clearly show scale-dependent PDF, with the typical increase of the tails as the scale decreases, indicating the increasing presence of bursts of energy transfer~\citep{sorriso2015}, typical of intermittency.
%
%
%
%
%
%
The scaling behavior of the PDFs of $|\epsilon_{\Delta t}|$ and $PVI_{\Delta t}(t)$ can be described through standard models of the turbulent cascade. 
The distribution of $|\epsilon_{\Delta t}|$ appears to be a stretched exponential function~\citep{sorriso2015}
  \begin{center}
  \begin{equation}
  P(|\epsilon_{\Delta t}|)\sim\exp(-b|\epsilon_{\Delta t}|^c)\, , 
  \label{stretched}
  \end{equation}
  \end{center}
where $b$ is a parameter related with the most probable value of the distribution, and $c\equiv c({\Delta t})$ describes the shape of the curve. In particular, for $c=2$ a Gaussian PDF is obtained, while $c=1$ corresponds to an exponential distribution. Values $0<c<1$ are associated with distribution whose tails can be more and more approximated by power-laws as $c$ decreases. The stretched-exponential distribution can be interpreted in the framework of the extreme deviations theory (EDT) applied to the fragmentation process occurring in the turbulent cascade, and more specifically when the statistics is controlled by a small number of extremely intense events~\citep{edt}. As turbulence is intermittent, this condition applies when the scale decreases, and EDT can be invoked to describe the statistics. 
Moreover, the PDF tail's ``flatness'' can be quantitatively represented through the parameter $c$, so that smaller $c$ corresponds to higher-tailed distributions and then to a higher probability of occurrence of extreme intense events. Thus, this is the most relevant scale-dependent parameter to describe the intermittency of the system.

\begin{figure}
  \begin{center}
  \includegraphics[width=\columnwidth]{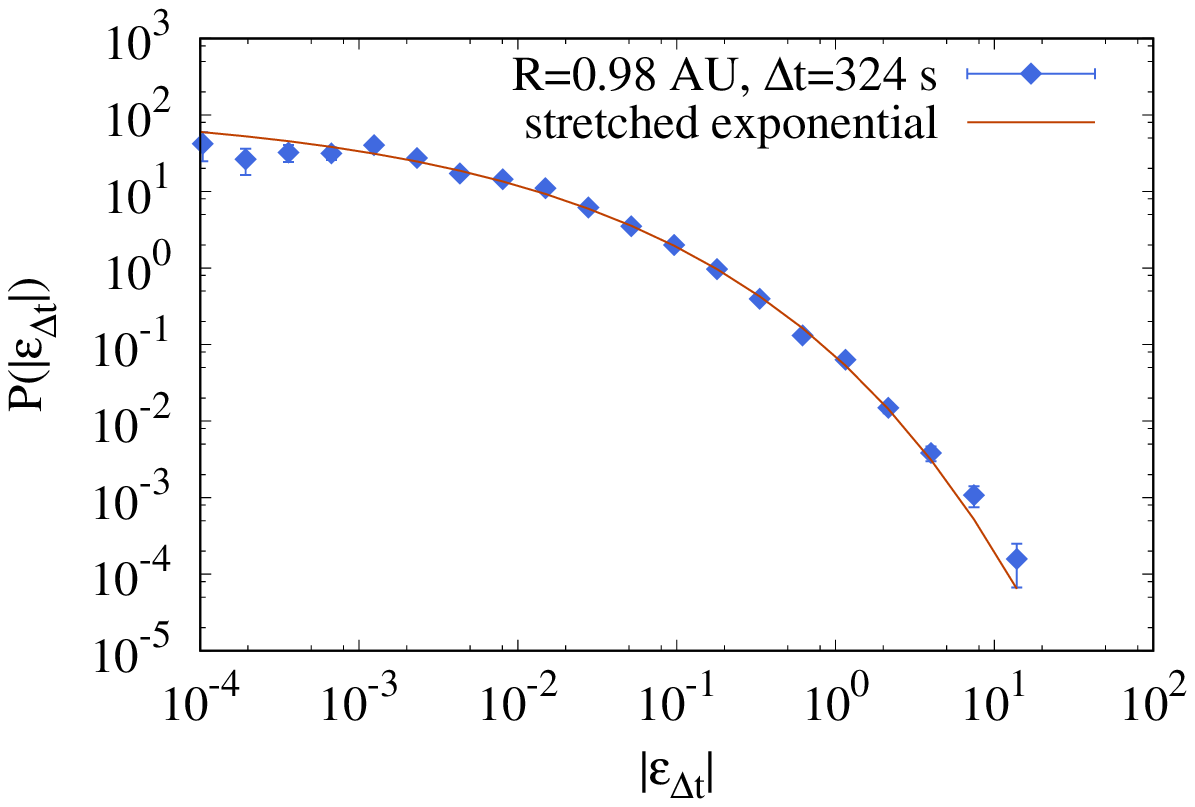}
  \includegraphics[width=\columnwidth]{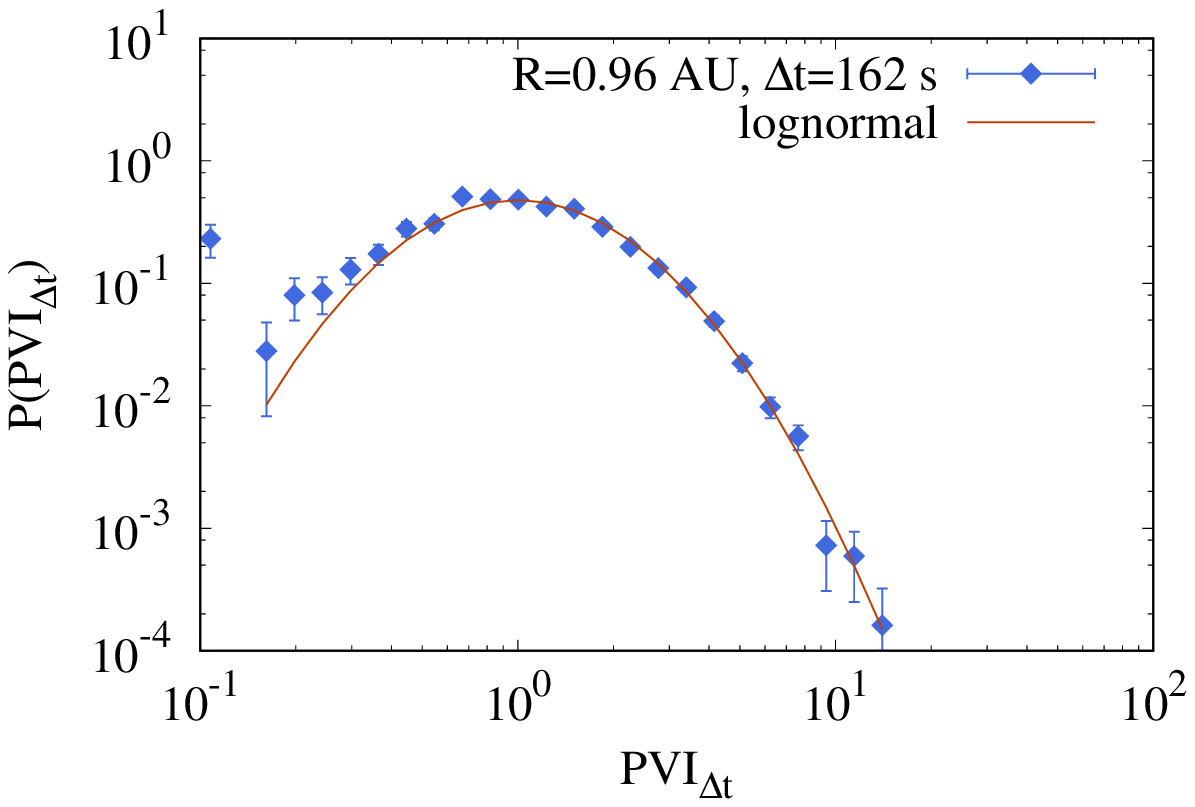}
  \end{center}
  \caption{Examples of fitted PDF of $|\epsilon_{\Delta t}(t)|$ for the fast wind interval of DAY 22 ({\it top panel}) 
           and $PVI_{\Delta t}(t)$ for the fast wind interval of DOY 32 ({\it bottom panel}) at the smallest time scale ${\Delta t}=81$ sec.
	   The error bars are obtained by propagating the standard Poisson statistical uncertainity of each bin count.}
  \label{Fig:fit}
  \end{figure}

\begin{figure}
  \begin{center}
  \includegraphics[width=0.8\columnwidth]{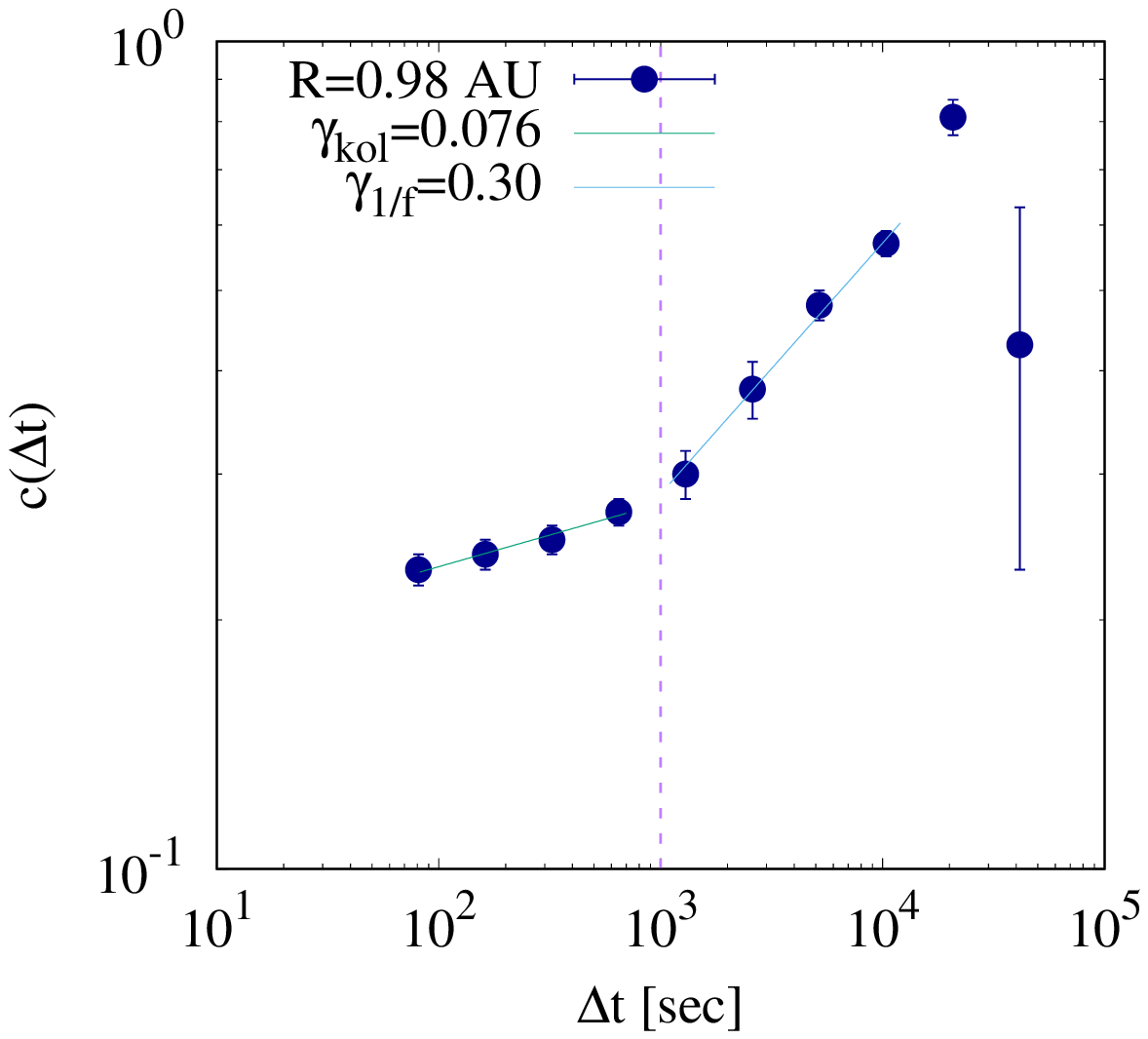}
  \includegraphics[width=0.8\columnwidth]{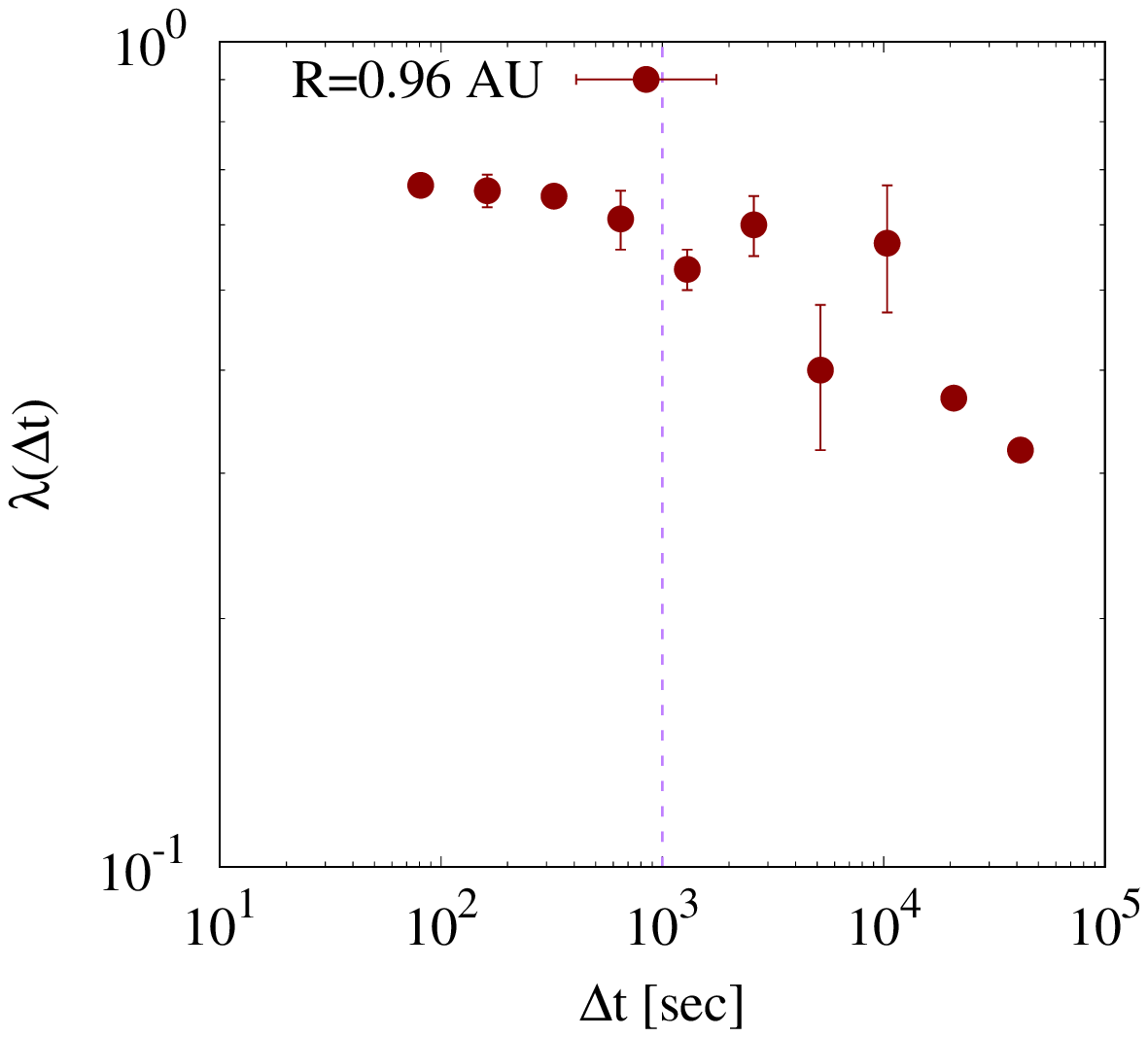}
  \end{center}
  \caption{Examples of scaling of the fitting parameters $c(\Delta t)$ for the fast wind interval of DOY 22 ({\it top panel}) 
	   and $\lambda(\Delta t)$ for the fast wind interval of DOY 32 ({\it bottom panel}), 
           obtained from the fit of the PDFs of $|\epsilon_{\Delta t}(t)|$ and $PVI_{\Delta t}(t)$, respectively. 
           The {\it vertical dashed line} indicates the turbulence correlation scale for those intervals~\citep{bruno2009}.
	   For the parameter $c$, two power-law fits are also superposed. The scaling exponent 
	   is $\gamma_{kol}=0.076\pm0.01$ in the inertial range (corresponding to $N_{c}\simeq 12$ steps 
           in the intermittent cascade) and $\gamma_{1/f}=0.30\pm0.02$ in the $1/f$ range (corresponding to 
           only $N_{c}\simeq 3$ steps in the cascade, if any exists).
	   In both panels, error bars represent the uncertainity of the parameters obtained from the $\chi^2$ minimization 
	   of the fitting procedure.}
  \label{Fig:tau}

  \end{figure}

The fit of $P(|\epsilon_{\Delta t}|)$ at one time scale for the interval of DOY 22 is displayed in the top panel of Figure~\ref{Fig:fit}. Similar fitting quality is achieved for all scales and all datasets. For each data interval, it is thus possible to describe the scale dependence of $P(|\epsilon_{\Delta t}|)$ through $c(\Delta t)$. The top panel of Figure~\ref{Fig:tau} shows one example of scaling of $c(\Delta t)$ for the same interval. In the same plot, the vertical bar indicates the turbulence correlation time estimated for this interval, $\tau_c\sim 1000$ sec~\citep{bruno2009}. Note that time scales smaller than the correlation time lye in the inertial range, while larger time scales are often associated with the $1/f$ spectral region~\citep{living}. A double power-law scaling $c(\Delta t)\sim{\Delta t}^\gamma$ is evident. Indeed, the break occurs near the correlation time, so that the scaling exponent $\gamma$ is different for the inertial range and for the $1/f$ range. Power-laws are often associated with the presence of correlations between bursts, indicating the presence of a non-stochastic process. The values found from power-law fits are $\gamma_{kol}=0.076\pm0.01$ and $\gamma_{1/f}=0.30\pm0.02$. According to EDT, $\gamma$ may be inversely proportional to the typical number of fragmentation steps $N_{c}$ occurring during the cascade. Therefore, larger $\gamma$ are associated with a smaller number of steps in the cascade. For the example given here $N_{c}\simeq 12$ in the inertial range, while $N_{c}\simeq 3$ in the $1/f$ range. 
Although the EDT model might not be fully adapted to describe solar wind turbulence, it can allow us to estimate the properties of intermittency based on the dissipation, at variance with standard approaches based on the field increments (PVI being an example of the latter). Moreover, the appearance of a power-law in the scaling of the parameter $c$ in the inertial range is indicative that $|\epsilon_{\Delta t}(t)|$ is a suitable variable for the description of the turbulent cascade.

On the other hand, as shown in the bottom panel of Figure~\ref{Fig:fit}, the PVI distribution can be well described by a log-normal function:
  \begin{center}
  \begin{equation}
  P(PVI_{\Delta t}) = \frac{1}{(\sqrt{2\pi}\lambda PVI_{\Delta t})}e^{-log(PVI_{\Delta t}-\langle PVI_{\Delta t}\rangle)^2/2\lambda^2}\, , 
  \label{lognormal}
  \end{equation}
  \end{center}
where $\langle PVI_{\Delta t}\rangle$ is the mean of PVI at the scale $\Delta t$ and $\lambda\equiv \lambda(\Delta t)$ is its scale-dependent variance, which determines the width of the distribution. 
The log-normal statistics of PVI can be naively understood in terms of the multiplicative process underlying the intermittent turbulent cascade~\citep{frisch}: at each position, for a given scale, the field increment is the result of all the previous fragmentation steps, which can be expressed in terms of multiplicative random factors. After a large number of fragmentations, the logarithm of the field increment will be the sum of the logarithms of randomly distributed multiplicative factors, so that the central limit theorem applies and the final $\log(PVI)$ value will obey Gaussian statistics. 
The fitting procedure provides, again, a quantitative estimation of the non-randomness of the fields, in terms of the presence of high tails. The parameter $\lambda(\Delta t)$ controls the width of the distribution, so that for larger $\lambda$, the distribution is broader, and the tails include the increasingly larger bursts of $PVI_{\Delta t}(t)$ arising at small scales because of intermittency. As for the stretched exponential parameter $c(\Delta t)$, the decrease of the parameter $\lambda(\Delta t)$ with the scale, shown in the bottom panel of Figure~\ref{Fig:tau} for DOY 32, suggests the non-self-similar nature of the fluctuations, although a clear power-law is not identified in this case. This confirms that a common mechanism, the nonlinear energy cascade, underlies the generation of the energy bursts in the turbulent field. 
Note that a similar scale-dependence of the distributions was obtained analysing the coarse-grained energy dissipation in MHD numerical simulations, and a proxy of dissipation, similar to PVI, in solar wind data~\citep{Zhdankin}.
%
%
%
%
%
%
%
%
%
%
%

Finally, the estimation of a proxy for the local energy transfer rate permits the study of the multifractal properties of the turbulent cascade, which are related to intermittency~\citep{frisch}. In the framework of the multifractal cascade models, the energy dissipation rate should be distributed in space as a multifractal object~\citep{pv}. Thus, the inhomogeneous character of the time series $\epsilon_{\Delta t}(t)$ highlighted by our analysis can be interpreted in terms of the multifractal properties of the field. It is therefore interesting to use this approach to characterize the degree of intermittency of the solar wind data, as opposed to the standard approach based on the scaling properties of velocity and magnetic field increments~\citep{burlaga1992,macek2006,macek2011}. 
In order to do so, we have evaluated the multifractal spectrum of the proxy (|$\epsilon_{\Delta t}(t)$|), estimated at the resolution scale $\Delta t=81$ sec, using a standard box-counting procedure (see for example the details given in~\citet{macek2011,sorriso2017}). The presence of singular structures ({\it e.g.} the bursts of local energy transfer rate) is revealed by the power-law scaling of the $q$-th order partition functions ($\chi_q(\delta t) \propto \delta t^{\tau_q}$) of a suitably-defined coarse-grained probability measure associated to the LET, for each scale $\delta t$. Note that $\delta t$ is the time scale over which the coarse graining is computed and is not related to the scale $\Delta t$ used for field increments computation.
The set of scaling exponents ($\tau_q$) describes the inhomogeneity of the singularity strength, and thus the multifractal properties of the field~\citep{frischparisi}. In particular, the exponents are expected to depend linearly on the order $q$ for mono-fractal objects, where only one singularity exponent is present. On the contrary, deviation from linearity indicates multifractality, i.e. a broader set of singularity exponents~\citep{grassberger}. 
Such deviation can be estimated using theoretical models, {\it e.g.} the p-model~\citep{pmodel}, which is the one adopted in this work. The p-model was originally developed for the description of the energy cascade in Navier-Stokes turbulence. It is a simple representation of the cascade in which the energy at one given position and scale is redistributed unevenly to two smaller scale (or ``daughter'') structures. The fraction of energy transferred at each step to each daughter structure is given by a cascade of multipliers randomly extracted from a binomial distribution, {\it i.e.} $p$ or $1 - p$, where $0\leq p \leq 0.5$ is the parameter that determines the characteristics of the cascade. In this simplistic view, the scaling exponents $\tau_q$ are directly related to the value of $p$ through $\tau_q=-log_2 \left[p^q+(1-p)^q\right]$~\citep{pmodel}. The parameter $p$ is thus a good quantitative measure of the deviation from self-similar (or fractal) scaling, {\it i.e.} of the degree of multifractality of the system. In particular, $p \simeq 0.5$ is an indication of mono-fractal fields, while smaller values are associated with greater multifractality.

\begin{figure}
  \begin{center}
  \vskip 12pt
  \includegraphics[width=0.7\columnwidth]{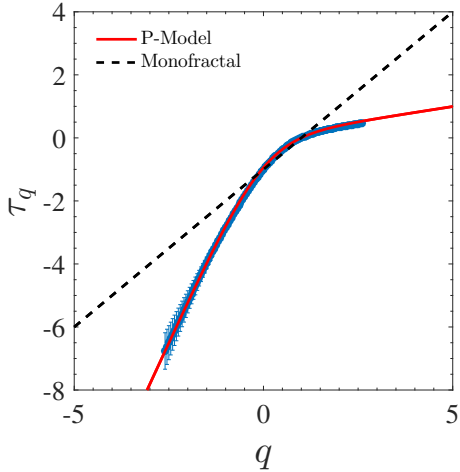}
  \end{center}
  \caption{Scaling exponents $\tau_q$ ({\it symbols}) 
	   for the {\it Helios 2} interval of DOY 22 with the p-model fit ({\it red line}). 
           For the interval in this example, $p=0.87$.}
  \label{Figtau}

  \end{figure}

For each solar wind $|\epsilon_{\Delta t}(t)|$ sample, the probability measures and their partition functions $\chi_q(\delta t)$ have been computed, for $q\in [-3, 3]$ with step $dq = 0.05$. 
The values of $\tau_q$ have then been evaluated by a fit of the partition functions to power-laws, in the range $\delta t\in[162,2000]$ seconds. Then, the exponents $\tau_q$ have been fitted to the p-model and the parameter $p$ has been estimated.
In Figure~\ref{Figtau} we show one example of the scaling exponents $\tau_q$ for DOY 22. A fit with the p-model is also indicated. The values of $p$ obtained for all intervals lie in the range $0.8\le p \le 0.9$, and are compatible with the usual strong intermittency parameters obtained using the fields increments~\citep{horbury1997,sorriso2017}.
Thus, the strong intermittent character of the field is well captured by such a multifractal analysis, based on a simple multiplicative model.
%
%
%
%
%
%
%
%
%
%
%
%
\section{Conditioned analysis of temperature}
\label{Section:Conditioned}
\begin{figure}
  \begin{center}
  \includegraphics[width=\columnwidth]{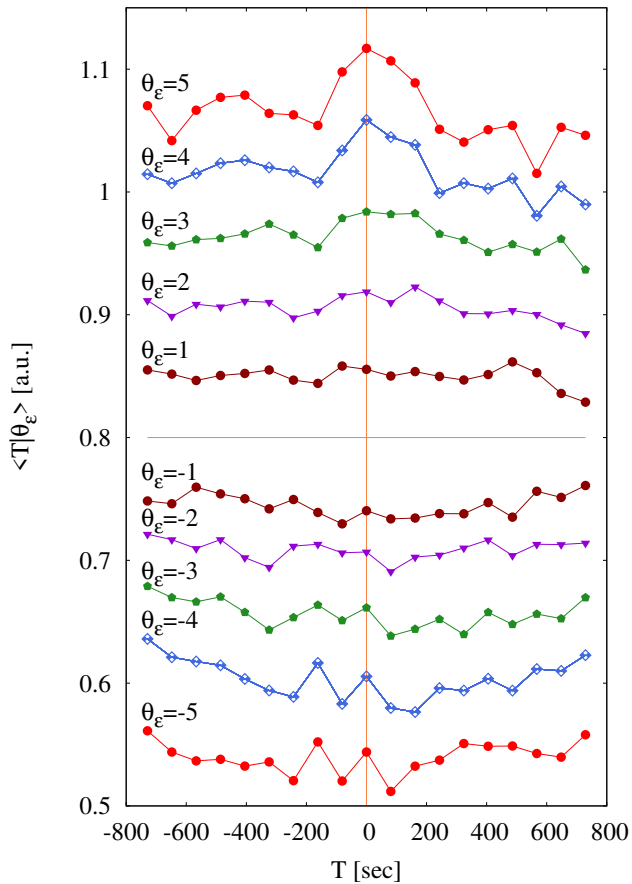}
  \end{center}
  \caption{Conditionally averaged normalized temperature $\langle T_p | \theta_{\epsilon} \rangle$ 
           as a function of the distance from the structure center, for different positive 
           and negative values of the threshold. All curves are arbitrarily vertically shifted for clarity.}
  \label{Fig:osman-eps}
  \end{figure}

\begin{figure}
  \begin{center}
  \includegraphics[width=\columnwidth]{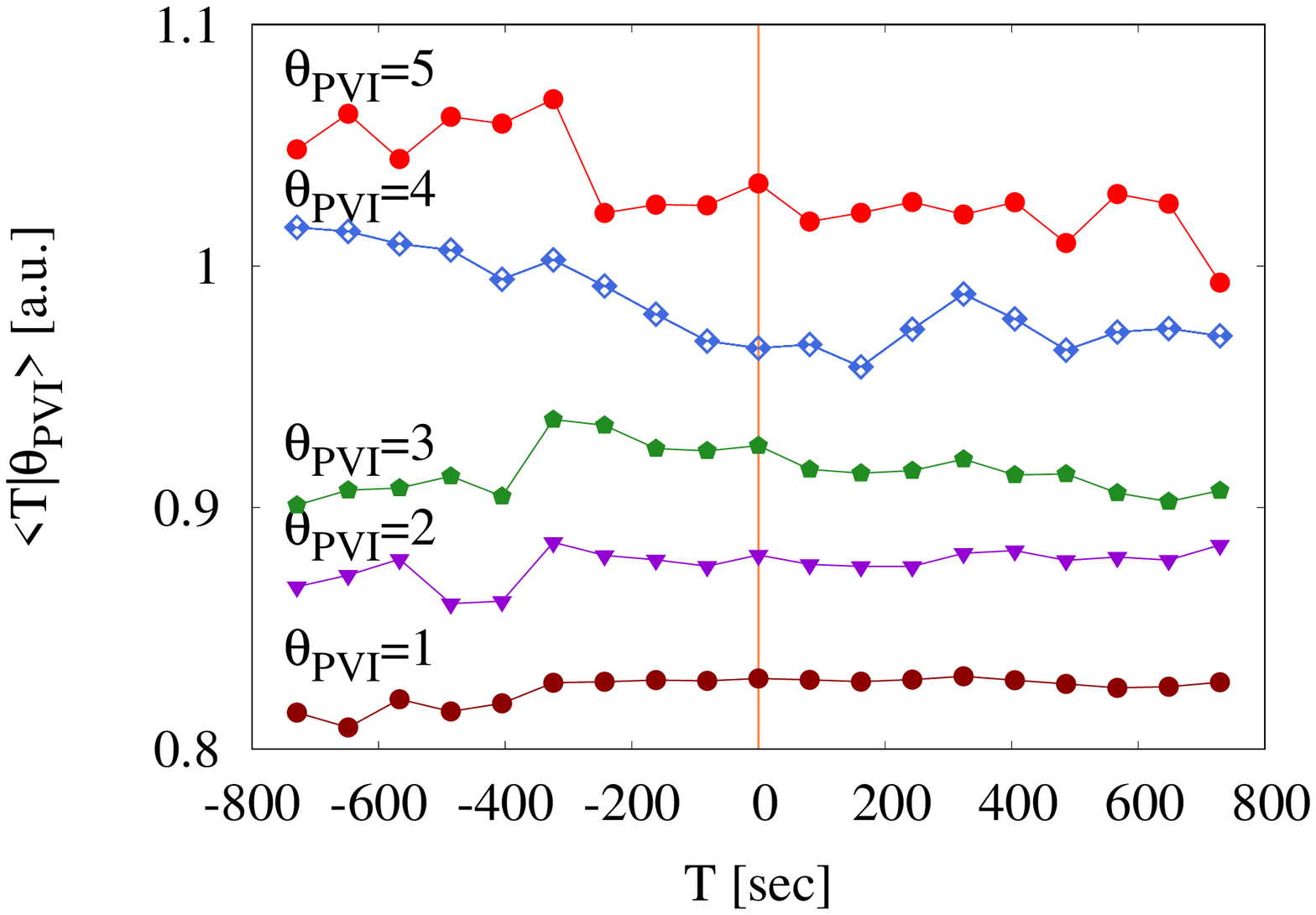}
  \end{center}
  \caption{Conditionally averaged normalized temperature $\langle T_p | \theta_{PVI} \rangle$ 
           as a function of the distance form the structure center, for different positive 
           values of the threshold. All curves are arbitrarily vertically shifted for clarity.
	  }
  \label{Fig:osman-pvi}
  \end{figure}

After describing the general statistical properties of the LET, it is interesting to point out its relationship with solar wind temperature. To this aim, the same analysis carried out by~\citet{osman} has been performed on {\it Helios 2} data using both the PVI and the LET at the resolution scale $\Delta t=81$ sec, for the localization of the small-scale structures. In this case, all 11 intervals were analysed together, in order to increase the statistical significance of the conditioning procedure. For this reason, for each sample the temperature time series was previously normalized to its mean value, allowing the comparison between different heliocentric distances and different wind type.
Upon identification of five incremental thresholds of the PVI values, and ten (five positive, five negative) of the LET, the conditioned average proton temperature has been estimated as a function of the distance $D$ from each PVI or LET structure, {\it i.e.} where their values overcome a given threshold, $\theta_{PVI}$ and $\theta_{\epsilon}$ respectively. 
This procedure provides the secular temperature profile around structures, $\langle T_p | \theta_{PVI} \rangle$ and $\langle T_p | \theta_{\epsilon} \rangle$. 
Figure~\ref{Fig:osman-eps} shows these temperature profiles as a function of the distance $D$ from the structures, conditioned to the indicated thresholds $\theta_{\epsilon}$ (coded in different colors and symbols). Note that all the curves have been arbitrarily shifted vertically for clarity.
While the low-threshold curve ($\theta_\epsilon$=1) is approximately constant (no local temperature increase), there is a striking evidence of temperature increase localized near the energetic LET structures, which is more evident as the conditioning value of $\epsilon_{\Delta t}$ is increased. In particular, for ($\theta_\epsilon$=5) the amplitude of the central peak indicates approximately 8\% higher temperature where the energy transfer rate is higher.
Such evidence results from the robust presence of hotter plasma near the structures, whereas the random fluctuations of the temperature are statistically canceled out far from these. There is also the appearance of a typical size of the higher-temperature site, which is of approximately 160 seconds around the structure. This might depend on the scale under study, so that a deeper analysis is left to an investigation in progress, based on higher resolution data. 
For the data associated with negative energy flux (lower part of the panel), the temperature profile has an evident threshold-dependent, incremental decrease approaching the structure, suggesting that whenever the energy flows from smaller to larger scale, the plasma is heated less than on average. This effect is less localized than the possible heating observed at positive $\epsilon_{\Delta t}(t)$, and may be similar to the effect observed by~\citet{osman} for low PVI data. 
Therefore, it is evident that there is a strong localization of higher plasma temperature near the sites of larger energy flux towards the small scales, possibly associated with local plasma heating, while the times with larger, negative energy flux are associated to colder plasma.
For comparison, the same analysis was carried out using PVI, {\it i.e.} repeating the~\citet{osman} procedure, and it is depicted in Figure~\ref{Fig:osman-pvi}. In this case, indication of higher temperature at the PVI structures is strongly reduced with respect to LET. 

The discrepancy between LET and PVI conditioning is related to the difference, already observed in Figure~\ref{Fig:scatter}, between the two proxies for large values, which are the most relevant for this analysis. The better performance of the LET shows that it is a more sensitive proxy, able to highlight the possible turbulent heating properties even when using limited size dataset.
%
%
%
%
%
%
%
%
%
%
%
%
%
%
\section{Conclusions}
\label{Section:Conclusions}
Unveiling the connection between the processes occurring at fluid scales and at kinetic scales ranges is important to understand how weakly collisional space plasmas dissipate the energy cascading from large to small scales. In particular, the solar wind represents an example of a collisionless plasma with a clear indication of heating due to turbulent energy dissipation. At the same time, it has the important advantage of the availability of in-situ measurements, at variance with other astrophysical plasmas. 
In this article we have presented one possible tool for the identification of the local transfer of turbulent energy across scales, the LET $\epsilon_{\Delta t}(t)$, based on the third-order moment scaling law for MHD turbulence~\citep{pp98}. This proxy differs from the usual tools such as LIM and PVI, as it includes, besides the direct kinetic and magnetic energy contributions, also cross-terms representing for example the cross-helicity contribution. The statistical analysis of the proxy provides insight on the scaling properties of the turbulent cascade, consistent with the standard turbulence analysis of solar wind plasmas. We have used {\it Helios 2} data to describe the properties of LET as compared to the PVI. Good correlations are found between PVI and LET, confirming that both indicators are suitable for the description of the turbulent cascade of energy. However, for large energy flux important differences arise. Moreover, the signed variable $\epsilon_{\Delta t}(t)$ carries information about the possible direction of cross-scale energy flow, which is hidden in the positive-defined variables LIM and PVI. This could be useful for better understanding the coupling mechanisms occurring near the MHD break scale. The study of this particular aspect is being addressed in a different work. 
The LET has also been studied through multifractal analysis based on the dissipation, rather than on the field increments. Such alternative analysis has confirmed the highly intermittent character of solar wind MHD turbulence.
Finally, a convincing correspondence between times of enhanced energy transfer rate and local temperature increase has been clearly demonstrated, indicating that LET is a useful tool for the identification of regions of interest for the study of turbulent energy dissipation. 

Because of its ability to track the link between the two range of scales, the LET could be useful for interpreting data from numerical simulations of the Vlasov-Maxwell system for the description of kinetic processes in collisionless plasmas. Similarly, it could have important implications for the analysis and interpretation of data from space missions providing high-resolution plasma measurements, such as MMS~\citep{mms} and the ESA candidate mission THOR~\citep{thor}, but also for the forthcoming Parker Solar Probe~\citep{probe} and Solar Orbiter~\citep{orbiter}.

\acknowledgments{
SP acknowledges support by the Agenzia Spaziale Italiana under the contract ASI-INAF 2015-039-R.O ``Missione M4 di ESA: Partecipazione Italiana alla fase di assessment della missione THOR''.
RM acknowledges financial support from the program PALSE (Programme Avenir Lyon Saint-Etienne) of the University of Lyon, in the frame of the program Investissements d’Avenir (No. ANR-11-IDEX-0007).
The authors declare they have no conflict of interest arising from the above funding.
}


\end{document}